\documentclass{article}
\usepackage{spconf,graphicx}
\usepackage{amsmath,amssymb}
\usepackage{url,lipsum,cite}
\usepackage{multicol,multirow}
\usepackage{xpatch,xcolor}
\usepackage{tabularx}
\usepackage{makecell, booktabs}
\newcolumntype{Y}{>{\centering\arraybackslash}X}
\usepackage{hyperref}
\mathchardef\mhyphen="2D

\hypersetup{
    colorlinks=true,
    linkcolor=purple,
    filecolor=magenta,      
    urlcolor=purple,
}

\newcommand{\newpara}[1]{\vspace{8pt}\noindent\textbf{#1}}
\newcommand\blfootnote[1]{%
  \begingroup
  
  \addtocounter{footnote}{-1}%
  \endgroup
}
\usepackage[symbol]{footmisc}

\title{RawBoost: A Raw Data Boosting and Augmentation Method\\ applied to Automatic Speaker Verification Anti-Spoofing}
\name{ Hemlata Tak, Madhu Kamble, Jose Patino, Massimiliano Todisco and Nicholas Evans
\thanks{The first author is supported by the VoicePersonae project funded by the French Agence Nationale de la Recherche (ANR) and the Japan Science and Technology Agency (JST).}}
\address{EURECOM, Sophia Antipolis, France}
\begin{document}
\ninept
\maketitle

\begin{abstract}
This paper introduces RawBoost, a data boosting and augmentation method for the design of more reliable spoofing detection solutions which operate directly upon raw waveform inputs.  While RawBoost requires no additional data sources, e.g.\ noise recordings or impulse responses and is data, application and model agnostic, it is designed for telephony scenarios.  Based upon the combination of linear and non-linear convolutive noise, impulsive signal-dependent additive noise and stationary signal-independent additive noise, RawBoost models nuisance variability stemming from, e.g., encoding, transmission, microphones and amplifiers, and both linear and non-linear distortion.  Experiments performed using the ASVspoof 2021 logical access database show that RawBoost improves the performance of a state-of-the-art raw end-to-end baseline system by 27\% relative and is only outperformed by solutions that either depend on external data or that require additional intervention at the model level.

\end{abstract}


\section{Introduction}
\label{sec:intro}
The recent ASVspoof 2021 challenge~\cite{delgado2021_ASV_spoof} addressed the problem of spoofing or presentation attack detection (PAD) in a logical access (LA) scenario in which both bona fide and spoofed utterances are encoded and transmitted across telephony networks.  The task was to learn reliable detection solutions using only the training and development partitions of the ASVspoof 2019 LA datasets which are without such encoding and transmission effects.
There is hence an interest in data augmentation techniques to compensate for the lack of in-domain training and development data~\cite{delgado2021_ASV_spoof,yamagishi2021_ASV_spoof}.

Still with a broad range of text-to-speech (TTS) and voice conversion (VC) spoofing attacks, the challenge maintained the focus of previous editions upon the development of generalised countermeasures that perform reliably not only in the face of spoofing attacks generated with TTS and VC algorithms different to those seen in training and development data, but also unseen encoding and transmission conditions.
While numerous data augmentation solutions have been proposed, e.g.\  SpecAugment~\cite{park2019specaugment} and  SpecMix~\cite{kim2021specmix}, they are suitable only for spoofing detection models which operate on two-dimensional front-end representations. End-to-end (E2E) spoofing detection solutions which operate on raw waveforms rather than two-dimensional representations are now gaining popularity~\cite{jung2020improved,tak2021rawnet,ma2021rw,tak2021end,ge2021raw,chen2021ur,caceres2021biometric}. The usual data augmentation techniques are then incompatible; they cannot be applied directly to raw waveform inputs. We have hence explored data augmentation techniques that are compatible with our RawNet2~\cite{tak2021rawnet} and RawGAT-ST~\cite{tak2021end} systems.

We introduce RawBoost, a data boosting and augmentation technique that can be applied directly to raw audio.  The aim is to improve spoofing detection reliability in the face of nuisance variation stemming from unknown encoding and transmission conditions which typify the LA or telephony scenario.  RawBoost is based upon well-known signal processing techniques and is computationally inexpensive with regard to the cost of learning on augmented data.  Furthermore, unlike WavAugment~\cite{kharitonov2021data}, a popular approach to data augmentation through pitch modification, band reject filtering, time dropping or the addition of reverberation or noise, techniques which can all be applied to raw waveforms, RawBoost operates without the need for any additional data sources, e.g., noise recordings or impulse responses, and neither requires any intervention at the model level.
 
RawBoost is data, application and model agnostic.  While we report its application to improve spoofing detection performance, it might have application to other related classification tasks where similar nuisance variability is expected, e.g., automatic speaker verification or automatic speech recognition.

\section{Data augmentation}
Data augmentation (DA) is commonly applied in many machine learning tasks to generate new samples from a source database, here utterances, to augment the pool of data available for training. The use of additional augmented data which exhibits variability not contained in the source data can help to reduce overfitting and bias, and hence to improve classification performance. 
Nowadays, DA is an integral component of modern machine learning pipelines and has been applied successfully in a host of different machine learning fields, such as image processing~\cite{wang2017effectiveness}, speech recognition~\cite{ko2015audio,vu2019audio} and speaker verification~\cite{zhang2018analysis}.  Recent work has also demonstrated its use in anti-spoofing~\cite{chen2020generalization,zhao2020replay,das2021data,das2021known,zhang2021empirical,chen2021ur,caceres2021biometric}. A number of approaches to DA have been proposed in the literature, e.g., random cropping, rotation and mirroring for image-related tasks~\cite{krizhevsky2012imagenet}; speed perturbation, pitch shifting, time stretching, random frequency filtering, reverberation, text-to-speech data augmentation and vocal tract length transformations for speech-related tasks~\cite{jaitly2013vocal,cui2015data,kharitonov2021data}. 

Knowing that evaluation data would contain both bona fide and spoofed utterances treated with a variety of unknown codecs, participants of the ASVspoof 2021 LA challenge used, e.g., speed perturbation~\cite{ko2015audio},  SpecAugment~\cite{park2019specaugment} and codec augmentation~\cite{vu2019audio} to help improve performance. SpecAugment, a form of spectral-domain augmentation, is applied to mask random intervals or bands of the spectrum and/or temporal frames during training but cannot be applied easily at the waveform level. Interest in raw E2E techniques for spoofing detection is currently growing~\cite{dinkel2017end,tak2021rawnet,ma2021rw,tak2021end,ge2021raw,chen2021ur,caceres2021biometric}.
There is hence a need for DA techniques that account for the variability expected in LA or telephony scenarios and, in particular, techniques that can also be applied at the raw waveform level.

\section{RawBoost Data Boosting and Augmentation}

\begin{figure*}[!h]
 \centering
 \includegraphics[width=18cm]{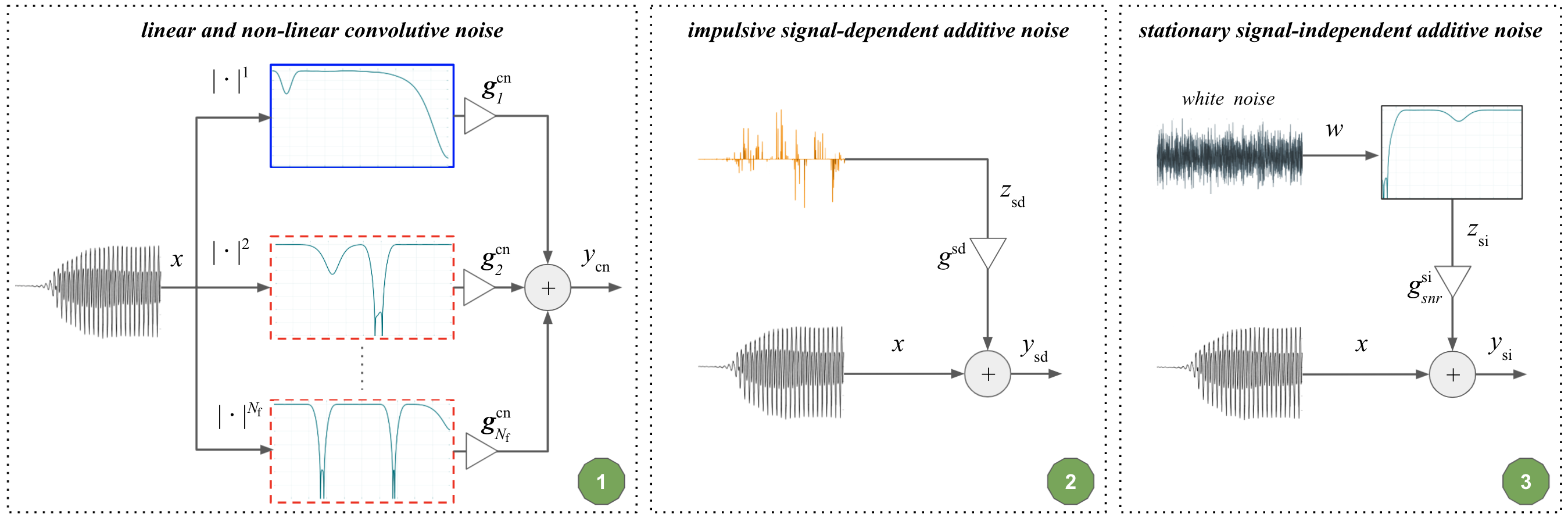}
  \caption{Proposed RawBoost data augmentation framework including: (1)~linear and non-linear convolutive noise; (2)~impulsive signal-dependent additive noise; (3)~stationary signal-independent additive noise. In (1), the profile in each rectangular box shows the frequency response for first harmonic (linear, solid blue box) and higher order harmonics (non-linear, dashed red boxes).}
\label{Fig:rawboost}
\end{figure*}

RawBoost\footnote{\url{https://github.com/TakHemlata/RawBoost-antispoofing}} is a data boosting and augmentation method which operates at the raw waveform level. Signal boosting approaches in machine learning have been gaining ground recently. Data boosting can encode prior knowledge about data or task-specific invariances, act as a regulariser to prevent overfitting, and can improve model robustness~\cite{guo2004boosting}. 
RawBoost employs established linear and non-linear signal processing techniques to boost or distort a set of utterances in a training dataset and/or augment a dataset with additional training utterances. RawBoost is illustrated in Fig.~\ref{Fig:rawboost} and comprises the three independent processes described below.

\subsection{Linear and non-linear convolutive noise}

Any channel involving some form of encoding, compression-decompression and transmission introduces stationary convolutive distortion. Most such channels will also introduce non-linear disturbances which are themselves also subject to stationary convolutive distortion, but of different characteristics (see~\cite{wang2020asvspoof}, Fig.~6).  In order to improve robustness to such nuisance variation, we have explored the combination of multi-band filtering and Hammerstein systems~\cite{kibangou2006wiener}. Hammerstein systems are proven, popular models of non-linear, dynamic systems in which non-linear static and linear dynamic subsystems are separated into different orders~\cite{kibangou2006wiener}. While Hammerstein models estimate multi-band filters from the response of non-linear systems, here we use the same idea to generate signal distortions.

\begin{figure}[t]
\centering
\includegraphics[width=8.6cm]{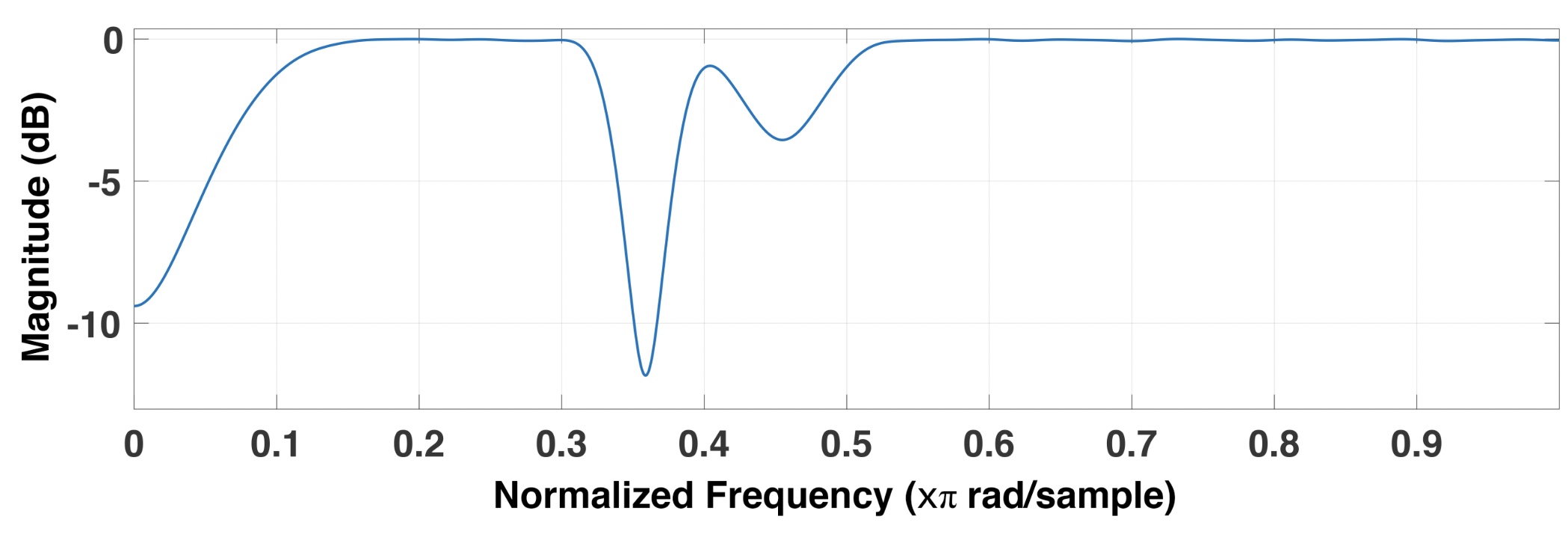}
\caption{Magnitude response of a multi-band filter with $N_{\textrm{notch}} = 3$ notch filters centered at normalised frequencies of 0.01, 0.35 and 0.45, bandwidths 0.06, 0.03 and 0.02 and number of filter coefficients 30, 94 and 52.} 
\label{fig:freqmask1}
\end{figure}

\newpara{Multi-band filters} are designed to generate convolutive noise using time domain notch filtering.
They are applied to a single utterance at a time and with a set of $N_{\textrm{notch}}$ notch filters, each with randomly chosen center frequencies $f_c$ and filter widths $\Delta f$. A single finite impulse response (FIR) filter with randomly chosen gain value $g_j^{\textrm{cn}}$ is then defined using a window-based filter design method~\cite{oppenheim2011discrete}, resulting in a filter with the desired frequency response using a randomly chosen number of filter coefficients~$N_{\textrm{fir}}$.  
The higher the number of coefficients, the more abrupt the frequency response; filters with fewer coefficients will exhibit passband ripple or distortion in addition to smoother cut-in and cut-off responses.  An example filter frequency response is illustrated in Fig.~\ref{fig:freqmask1}.  It has $N_{\textrm{notch}}=3$ notch filters, each with different center frequencies, stop-band widths and number of filter coefficients. 

\newpara{Hammerstein systems} generate higher-order harmonics whereby a component $f_0$ in the input to a non-linear system is supplemented at the output by $N_{\textrm{f}}-1$ new components at $2f_0, 3f_0,...,N_\textrm{f}f_0$, leading to non-linear harmonic distortion. The frequency and amplitude of each higher-order harmonic are dependent upon those of the original component and the characteristics of the non-linear system. 

Convolutive noise $y_{\textrm{cn}}$, denoted \textcircled{1} in Fig.~\ref{Fig:rawboost}, is generated according to:

\begin{equation}
    \begin{aligned}
    y_{\textrm{cn}}[n] = \sum_{j=1}^{N_{\textrm{f}}} g_j^{\textrm{cn}}\sum_{i=0}^{N_{\textrm{fir}_{j}}} b_{i_j} \cdot x^j[n-i]
    \end{aligned}
    \label{eq: 3}
\end{equation}

\noindent where $x\in[-1,1]^{l \times 1}$ denotes a raw waveform of $l$ samples, $j\in[1,N_\textrm{f}]$ is the order of the (non-)linearity ($N_\textrm{f}=1$ refers to the filter applied to the linear component $x$), $b_{i_j}$ denotes the coefficients of the $j^{\textrm{th}}$ multi-band filter.

\subsection{Impulsive signal-dependent additive noise}
Impulsive signal-dependent noise is commonly introduced through data-acquisition, resulting from, e.g., clipping, non-optimal device operation (microphones and amplifiers), synchronization and overflow issues, or as a result of insufficient computational power.  It is typically orders of magnitude lower in amplitude than signal-independent noise~\cite{kok1998multirate}.
We model such nuisance variability as non-stationary impulsive disturbances (see \textcircled{2} in Fig.~\ref{Fig:rawboost}) consisting of instantaneous or impulse-like amplitude variations. The disturbance $z_{\textrm{sd}}$ is applied to a maximum of $P\leq l$ 
uniformly distributed
samples $\{p_1, p_2,...,p_P\}$ in $x$ to obtain ${y}_{\textrm{sd}}$ according to:
\begin{equation}
{y}_{\textrm{sd}}[n]= x[n] + z_{\textrm{sd}}[n]
    \label{eq: 5}
\end{equation}
\noindent where
\begin{equation}
    z_{\textrm{sd}}[n]=
    \begin{cases}
    g^{\textrm{sd}} \cdot D_R\{-1,1\}[n]  \cdot x[n],& \text{\textit{if} } n = \{p_1, p_2,...,p_P\}\\
    0,              & \text{\textrm{otherwise}}
    \end{cases}
     \label{eq: 4}
\end{equation}
\noindent is a \textit{signal-dependent additive noise} component, $g^{\textrm{sd}} > 0$ is a simple gain parameter and where $D_R\{-1,1\}[n]$ denotes $P$ values randomly chosen from the distribution:
\begin{equation}
    f_R(r)= 
    \begin{cases}
  -log(r),&  0<r\leq1\\
    -log(-r),              & -1\leq r<0
    \end{cases}
\end{equation}
\noindent For convenience, the maximum number of samples $P$ is chosen relatively as $P_{\textrm{rel}}=P/l$.

\subsection{Stationary signal-independent additive noise}
The use of signal-independent, additive noise is one of the most popular forms of data augmentation and has been applied in a wide variety of applications, including speech recognition~\cite{yin2015noisy}, speaker recognition~\cite{huh2020augmentation}, speech emotion recognition~\cite{tiwari2020multi}, as well as audio forgery~\cite{yang2014additive} and spoofing detection~\cite{das2021data}.  Signal-independent additive noise can result from loose or poorly joined cable connections, transmission channels effects, electromagnetic interference or thermal noise. In contrast to impulsive noise, a stationary white noise~$w$ (see \textcircled{3} in Figure~\ref{Fig:rawboost}) is coloured using a FIR filter designed in the same way as described in Section 3.1, before being added to the entire utterance:
\begin{equation}
    y_{\textrm{si}}[n] = x[n] + g^{\textrm{si}}_{snr} \cdot z_{\textrm{si}}[n]
    \label{eq: 6}
\end{equation}

\noindent where
\begin{equation}
    g^{\textrm{si}}_{snr} = \frac{10^{\frac{SNR}{20}}}{\lVert{z_{\textrm{si}}}\rVert^2 \cdot \lVert{x}\rVert^2} 
     \label{eq: 7}
\end{equation}
\noindent is a gain parameter corresponding to a randomly chosen SNR and where $z_{\textrm{si}}$ is the result of white noise $w$ coloured by the FIR filter.

\section{Experiments and results}
\label{sec:ExpAndResult}
Described in this section are the database, evaluation metric, baseline system and our results.  
\subsection{Dataset, protocols and metrics}
\label{sec:asv2021_challenge}
The ASVspoof 2021 logical access (LA) task focuses on the development of spoofing and deepfake detection solutions that are robust to encoding and transmission channel variability.  Spoofed speech data are generated with the same text-to-speech (TTS), voice conversion (VC) and hybrid algorithms (VC with TTS-generated inputs) used for the 2019 challenge.  In contrast to the ASVspoof 2019 LA training and development data, all ASVspoof 2021 evaluation data is transmitted across some form of communications network, e.g., a public switched telephone network (PSTN) or a voice over Internet Protocol (VoIP) network using one of a number of different, popular telephony codecs, e.g., A-law and G.722 codecs, though other, unknown or  unannounced codecs were also used (codec and other meta data was withheld from participants), giving the following conditions: \textbf{C1}: no encoding/transmission, \textbf{C2}: A-law, VoIP; \textbf{C3}: PSTN; \textbf{C4}: G.722, VoIP; \textbf{C5-C7}: unknown. As per the 2021 challenge rules, we used only the ASVspoof 2019 LA training and development partitions in optimising our spoofing countermeasure.  We used the default minimum normalised tandem detection cost function (t-DCF)~\cite{kinnunen-tDCF-TASLP} as a primary metric but also report results in terms of the pooled equal error rate (EER).

\begin{table}[!t]
\centering
\scriptsize
\renewcommand{\arraystretch}{1.4}
\setlength\tabcolsep{1.75pt}
\caption{RawBoost parameter values. Values within expressed ranges are selected at random (uniform distributions).}
\label{tab:param}
\begin{tabular}{|c|c|c|c|c|c|c|c|c|c|c|}
\hline 
Param&$N_{\textrm{notch}}$ &$N_{\textrm{fir}}$ &$N_{f}$ &$f_c$ &$\Delta f$ & $g^{\textrm{cn}}_{1}$ & $g^{\textrm{cn}}_{2 \mhyphen N_{f}}$ & $P_{\textrm{rel}}$&$g^{\textrm{sd}}$&$SNR$ \\
&&&&[Hz]&[Hz]&[dB]&[dB]&[\%]&&[dB]\\

\hline\hline
\textcircled{1} &5&[10,100]&5&[20,8k]&[100,1k]&[0,0]&[-5,-20]&-&-&-\\
\hline
\textcircled{2} &-&-&-&-&-&-&-&[0,10]&2&-\\
\hline
\textcircled{3} &5&[10,100]&1&[20,8k]&[100,1k]&-&-&-&-&[10,40]\\
\hline
\end{tabular}%

\end{table}

\subsection{Baseline}
\label{sec:baseline}
The baseline is an end-to-end RawNet2 system~\cite{tak2021rawnet}. It is among the best-performing single systems and all results are fully reproducible using open source software.\footnote{\url{https://github.com/asvspoof-challenge/2021/tree/main/LA/Baseline-RawNet2}} The same system was adopted as one of four baselines for the ASVspoof 2021 challenge~\cite{delgado2021_ASV_spoof,yamagishi2021_ASV_spoof}. The first sinc layer is initialised with a bank of 20 mel-scaled filters. Each filter has an impulse response of 1025 samples (64~ms duration) which is convolved with the raw waveform. The latter are truncated or concatenated to give segments of {~$4$}~seconds duration (64,600 samples). The sinc layer is followed by a residual network and a gated recurrent unit (GRU) to predict whether the input audio is bona fide or spoofed. We used the Adam optimiser with a mini-batch size of $128$ and a fixed learning rate of $0.0001$ and train for $100$ epochs. Full details of the baseline system are available in~\cite{tak2021rawnet}.

\vspace{-0.5cm}
\subsection{RawBoost configurations}
RawBoost parameters are generated according to the configuration options illustrated in~Table~\ref{tab:param} for each of the three techniques.  Values expressed within ranges are drawn from the corresponding uniform distributions. Each technique is applied alone as well as in different combinations and in both series and parallel.\footnote{Due to space limitation, only a selection of best results is reported.} For series combinations, the output of one technique is used as the input to the next.  For parallel combinations, an original input utterance is treated independently with each technique before the resulting distortions are combined. Output waveforms are normalised to prevent overflow. In our experiments, we used RawBoost to add nuisance variability on-the-fly to \emph{existing} training data, instead of to generate \emph{additional} data. Since the ASVspoof 2019 LA development data exhibits neither encoding nor transmission variability, and in order to respect properly the ASVspoof 2021 protocols and evaluation rules, we applied RawBoost also to the development data.  RawBoost parameters and ranges illustrated in Table~\ref{tab:param} were then selected based on the results of experimentation involving boosted and augmented training and development data only.  

\begin{table*}[!t]
    \centering
     \caption{ASVspoof 2021 LA RawNet2 results in terms of min t-DCF for each codec, pooled min~t-DCF (P1) and pooled EER (P2).} 

      \renewcommand{\arraystretch}{0.92}
    \small
    \setlength\tabcolsep{2.5pt}
    \begin{tabular}{|c|c||c|c|c|c|c|c|c||c|c|}
      \hline
      Augmentation&Method&C1&C2&C3&C4&C5&	C6&C7& P1 & P2 	\\ 

     \hline \hline
     none&-& 0.4629&	0.5594&	0.7886	&0.4954	&0.5582	&0.6774&	0.5727&	0.4257&9.50\\
     
     \hline\hline

       &(1) linear and non-linear convolutive noise &0.4531&	0.5077&	0.6160&	0.4731	&0.5019&	0.5819	&0.5317&	0.3527&7.22\\
        \cline{2-11}
        &(2) impulsive signal dependent noise &0.4373&	0.5015	&0.5041	&0.4751	&0.4920&	0.5385&	0.5099&	0.3260&6.09\\
         \cline{2-11}
        &(3) stationary signal independent noise &0.4544	&0.5094	&0.5349	&0.4811	&0.5036	&0.5289&	0.4964&	0.3372&7.85\\
          \cline{2-11}
         &\textbf{\textit{series}: (1)+(2)}&\textbf{0.4449}	&\textbf{0.4806}&	\textbf{0.5046}	&\textbf{0.4635}	&\textbf{0.4616}	&\textbf{0.5025}&	\textbf{0.4776}&	\textbf{0.3099}&\textbf{5.31}\\
         \cline{2-11}
         RawBoost&\textit{parallel}: (1)+(2)&0.4471&    0.5094  &  0.5507    &0.4724    &0.5032&    0.5585   & 0.5243&    0.3261&5.57\\
         \cline{2-11}
           &\textit{series}: (1)+(3)&0.4569   & 0.5203&    0.5576 &   0.4765    &0.5057    &0.5442    &0.5134   & 0.3361&6.27\\
          \cline{2-11}

           &\textit{series}: (2)+(3)&0.4640&    0.5056 &   0.5100 &   0.4910 &   0.5060   & 0.5240  &  0.5171  &  0.3329&6.58\\
          \cline{2-11}

         &\textit{series}: (1)+(2)+(3)&0.4437&    0.4910   & 0.4986&    0.4576   & 0.4937    &0.5037 &   0.4858    &0.3192&5.39\\

         \hline\hline
         
         &(1) time-drop &0.4582	&0.5049	&0.5133	&0.4598	&0.5094	&0.5296	&0.4739	&0.3490&8.72\\
         \cline{2-11}
        &(2) band-reject &0.4763	&0.5417&	0.5912	&0.4957	&0.5387&	0.5628&	0.5174&	0.3692&8.86\\
        \cline{2-11}
         WavAugment &(3) additive-noise &0.5508&0.6721	&0.7014	&0.5531	&0.6649	&0.6549&	0.5660&	0.4819&13.38\\
        \cline{2-11}
         &\textit{series}: (1)+(2)+(3)&0.4652&    0.4897 &   0.5172    &0.4736    &0.4802    &0.5163   & 0.4990   & 0.3435&7.32\\
        \cline{2-11}
         &\textit{series}: (pitch)+(reverberation)+(1)+(3)&0.6130&	0.7013&	0.7351&	0.6138&	0.7153&	0.7229&	0.6307&	0.5414&15.66\\
         
         \hline\hline
        &(1) frequency-masking &0.4579	&0.5292	&0.7171	&0.4894&	0.5399&	0.6642&	0.5335&	0.4214&9.80\\
        \cline{2-11}
        SpecAugment&(2) time-masking &0.4581	&0.5049	&0.5134	&0.4598	&0.5094	&0.5295	&0.4739	&0.3491&8.72\\
         \cline{2-11}
         &\textit{series}: (1)+(2)&0.4668  &  0.4985&    0.5032 &   0.4927   & 0.4918    &0.5162   & 0.4822    &0.3418&8.25\\
         \hline
   \end{tabular}
   
    \label{tab:breakdown}
\end{table*}

\subsection{Results}
\label{ssec:result}
Results are illustrated in Table~\ref{tab:breakdown} for the baseline system (row 1), and for the same system trained using one of the three approaches to data augmentation: RawBoost; WavAugment; SpecAugment. For SpecAugment experiments,\footnote{SpecAugment is not applied to raw waveforms but at the filterbank output instead.  Results are included nonetheless for comparison.  Only augmentation techniques applied at the raw waveform level support learning of the filterbank layer using augmented data.} frequency (channel) masking is applied at the sinc filterbank level to mask random contiguous sinc channels during training. In each case, results are shown for separate augmentation techniques and a selection of combinations (column~2). Columns 3-9 show results for each evaluation condition (C1-C7). Columns 10 and 11 show the pooled min t-DCF (P1) and pooled EER (P2). All RawBoost DA strategies lead to better performance than the baseline for all 7 evaluation conditions. The baseline pooled min t-DCF of 0.4257 drops to 0.3527 when using linear and non-linear convolutive noise~(1), to 0.3260 using impulsive signal dependent additive noise~(2) and to 0.3372 using stationary signal-independent additive noise~(3). The best result is obtained using the RawBoost (1)+(2) system for which the min t-DCF is 0.3099 (27\% relative reduction over the baseline) and the EER is 5.31\% (44\% relative reduction). The addition of stationary noise, while beneficial on its own, does not lead to any improvements in performance when combined with other techniques.  This is not entirely surprising given that ASVspoof LA data does not contain any ambient noise.  The technique may yet prove beneficial for other tasks, e.g., the physical access (PA) scenario, that {\em do} contain ambient noise.

\begin{table}[!t]
	\centering
	\small

	\caption{A performance comparison for the ASVspoof 2021 evaluation partition in terms of pooled min t-DCF and pooled EER for different state-of-the-art single systems. }
 \setlength\tabcolsep{1pt}
  \renewcommand{\arraystretch}{0.98}
	\begin{tabular}{ *{5}{c}}
		\hline\hline
		 system &front-end& DA approach & min t-DCF&EER	\\
		 \hline
		 LCNN~\cite{tomilov2021stc}&Mel STFT&RS Mixup and FIR&0.2430&2.21\\
		 \hline
		 ResNet-L-LDE~\cite{chen21b_asvspoof}&LFB&Frequency masking&0.2720&3.68\\
		 \hline
		 \textbf{\textit{Ours}}:RawNet2&Raw&RawBoost (1)+(2)&0.3099&5.31\\
		 \hline
		   SE-ResNet18~\cite{kang2021crim}&LFCC&codecs&0.3129&6.62\\
		   \hline
		
		RawNet2~\cite{caceres2021biometric}&Raw&codecs& 0.3168&6.36\\
		 \hline
		  LCNN~\cite{das2021known}&CQT&codecs&0.3197&5.27\\
\hline	\hline
\end{tabular}
	\label{Tab:comparsion results}
	\vspace{-0.5cm}
\end{table}

\vspace{-0.2cm}
\subsection{Comparison to competing systems}
\label{ssec:comparison_results}
Illustrated in Table~\ref{Tab:comparsion results} is a comparison of RawBoost performance to that of competing systems reported in the literature.  To focus upon the benefits of data augmentation, the comparison is restricted to single systems.\footnote{While some ensemble systems outperform those considered here, they are substantially more complex and their inclusion would compound the difficulty in assessing {\em data augmentation}. Unlike the comparisons made in Table~\ref{tab:breakdown}, differences in Table~\ref{Tab:comparsion results} stem from differences in data augmentation as well as the underlying models/classifiers.} The RawNet2 system with (1)+(2) RawBoost DA gives the third best result.  Among the top three systems, only RawNet2 operates upon raw waveform inputs.  The ResNet-L-LDE system~\cite{chen21b_asvspoof}, which uses SpecAugment data augmentation in the form of frequency masking, uses external data contained within the MUSAN database. In contrast, RawBoost requires no such external data.
The top-performing LCNN system~\cite{tomilov2021stc} uses random square (RS) mixup~\cite{zhang2018mixup} and FIR filtering DA.  The FIR filtering approach aims to emulate the application of different telephony codecs and is conceptually similar to our use of FIR filtering.  While applied at the data level, RS mixup is accompanied with modifications at the model level (the loss function in~\cite{tomilov2021stc}).  RawBoost requires no such intervention at the model level. 
The remaining systems included in Table~\ref{Tab:comparsion results} all augment data using speech codecs.  RawBoost is competitive with all these approaches while not requiring the use of any additional codec implementations.

\section{Conclusions}
\label{sec:conclusion}
RawBoost can be used to boost or augment the pool of data available for training by generating new utterances which exhibit the variability expected in telephony scenarios.  
New raw waveforms are generated by perturbing a set of source utterances using linear and non-linear convolutive, and both impulsive and stationary additive noise. Our results show that RawBoost improves the performance of a raw end-to-end baseline spoofing detection solution by up to 27\% relative. RawBoost is also data, application and model agnostic; it operates upon an existing source database without the need for any additional external data, nor intervention at the model level. It might hence have application to other related audio classification tasks.

\clearpage
\bibliographystyle{IEEEbib}
\bibliography{refs,Hemlata_GAT,Data_aug}

\end{document}